# Thermal Rectification In Asymmetric Graphene Ribbons

Nuo Yang[1], Gang Zhang[2, a)] and Baowen Li[3,1, b)]

[1]*Department of Physics and Centre for Computational Science and Engineering, National University of Singapore, 117542 Singapore*

[2]*Institute of Microelectronics, A*STAR, 11 Science Park Road, Singapore Science Park II, Singapore 117685, Singapore*

[3]*NUS Graduate School for Integrative Sciences and Engineering, 117597 Singapore*

## ABSTRACT

In this paper, heat flux in graphene nano ribbons has been studied by using molecular dynamics simulations. It is found that the heat flux runs preferentially along the direction of decreasing width, which demonstrates significant thermal rectification effect in the asymmetric graphene ribbons. The dependence of rectification ratio on the vertex angle and the length are also discussed. Compared to the carbon nanotube based one-dimensional thermal rectifier, graphene nano ribbons have much higher rectification ratio even in large scale. Our results demonstrate that asymmetric graphene ribbon might be a promising structure for practical thermal (phononics) device.

[a)] Electronic mail: zhangg@ime.a-star.edu.sg

[b)] Electronic mail: phylibw@nus.edu.sg

The graphene is a two-dimensional crystal consisting of a single atomic layer of carbon. It has attracted immense interests recently, mostly because of its unusual electronic properties [1]. Graphene ribbons (GRs), the narrow layers of graphene are particularly interesting as promising elements in nanoelectronics. It has been shown that many of the electronic properties of GR can be modulated by its size and shape [2-5]. In addition, the thermal properties of graphene are also important both for fundamental understanding of the underlying physics in low-dimensional system and for applications. Superior thermal conductivity has been observed in graphene [6, 7], which has raised the exciting prospect of using graphene structures for thermal (phononics) devices. Traditionally, phonons are considered as heat carriers. Recently, it is found that phonons can be also used to carry and process information [8-17]. More importantly, different thermal (phononic) devices such as thermal rectifier, thermal transistor, thermal logic gate, and thermal memory, have been conceptualized. All these have not only made the control of heat flow possible, but also made it possible to use phonons to carry and process information.

Similar to electronic counterpart, the thermal rectifier also plays a vital role in phononics circuit [16]. Thermal rectifier from nano structures has been theoretically studied [14, 17] and also experimentally tested [18] in carbon nanotube (CNT) based systems. All the structures studied so far are basically one-dimensional or quasi-one-dimensional systems. A natural question comes promptly: Can we extend the thermal rectification characteristic to two-dimensional structure, such as in GR? In this Letter, we should investigate numerically (by using molecular dynamics) the direction dependent heat flux in asymmetric structural GRs, and discussed the impacts of GR shape and size on the rectification ratio. Our study may inspire experimentalists to develop GR based thermal (phononic) devices.

In our simulations, classical non-equilibrium molecular dynamics (MD) method is adopted. The potential energy is a Stillinger-Weber type potential for bonding interaction between carbon atoms, which include both two-body and three-body potential terms. This force field potential is developed from fitting experimental parameters for graphite by Abraham and Batra [19] and the accuracy of this potential is demonstrated in MD simulations in graphitic structures [20].

Figure 1 shows two different GR structures: trapezia shaped GR (TGR) and two rectangular GRs with different width (RGR). In order to establish a temperature gradient along the longitudinal direction, the GR is coupled with Nosé-Hoover heat baths on the two end layers. Atoms in the same layer have the same z coordinate. The atoms of 1st and *Nth* (N is the number of layers) layers are frozen corresponding to fixed boundary condition. The atoms of 2nd and N-1'th layers are coupled with Nosé-Hoover heat baths [21, 22], whose temperatures are $T_{top}$ and $T_{bottom}$ respectively. In this Letter, in order to avoid any artificial effect induced by the choice of heat baths, all the calculations are performed with the same Nosé-hoover heat bath parameter (thermostat response time) on the two ends.

The velocity Verlet algorithm is used to integrate the differential equations of motions. In general, the temperature, $T_{MD}$, is calculated from the kinetic energy of atoms according to the Boltzmann distribution [23]. It is worth pointing out that this approach is valid only at high temperature ($T>>T_D$, $T_D$ is the Debye temperature). When the system temperature is lower than the Debye temperature, it is necessary to apply a quantum correction to the MD calculated temperature. The difference between the MD calculated temperature and quantum corrected temperature depends on the Debye temperature of the system. Although some theoretical studies have been performed on the Debye temperature of graphene, the results have been controversial. For instance, from 1000 K to 2000 K are reported. [24-26] The concerns of the current paper are the heat flux and rectification effects (a related change of heat flux) which do not depend on the accurate value of temperature, instead they depend on the temperature difference. Therefore, we don't do the quantum correction to $T_{MD}$ in this work. The heat flux along the GR is defined as the energy transported along it in unit time. [17] MD simulations are performed long enough such that the system reaches a stationary state when the local heat flux is a constant. All results given here are obtained by averaging $4\times10^6$ time steps. The time step is 0.4 fs.

In this Letter, we set the temperature of top (the short end) as $T_{top} = T_0(1-\Delta)$ and that of bottom (the long end) as $T_{bottom} = T_0(1+\Delta)$, where $T_0$ is the average temperature, and $\Delta$ is the normalized temperature difference between the two ends. Therefore, the bottom of GR is at a higher temperature when $\Delta > 0$, and the top has a higher temperature when $\Delta < 0$. First, we study the heat flux in TGR and RGR with L=3.4 nm (30 layers), $W_{top}$ = 0.42 nm, and $W_{bot}$ = 4.2 nm which corresponds to θ=60º in TGR, and $T_0$=300 K. The heat flux $J$ versus temperature difference $\Delta$ (corresponds to the "I-V" curve of electric rectifier) is shown in Figure 2(a). When $\Delta > 0$, the heat flux increases steeply with $\Delta$; while in the region $\Delta < 0$, the heat flux is much smaller and changes a little with $\Delta$. That is, all the two types of GRs, TGR and RGR behave as a "good" thermal conductor under positive "thermal bias" and as a "poor" thermal conductor under negative "thermal bias". It suggests that the heat flux runs preferentially along the direction of decreasing width.

This is similar to the rectification phenomena observed in carbon nanocone (CNC) structures [17], which was explained well by match/mismatch of the phonon spectra between the atomic layers at the two ends. To show the rectification effect quantitatively, the *thermal rectification* is defined as,

$$R \equiv \frac{(J_+ - J_-)}{J_-} \times 100\%$$

where $J_+$ is the heat current from bottom to top corresponds to $\Delta > 0$ and $J_-$ is the heat current from top to bottom when $\Delta < 0$. Figure 2(b) shows the rectifications with different $\Delta$. The increase of $\Delta$ results in the increase of the rectification ratio akin to the characteristic in electric rectifier and coincides with previous results of thermal rectification models [13, 14, 17]. The rectification ratios of GRs are much higher than those in CNC and CNT junctions. With $|\Delta|=0.5$, the rectification of nanocone is 96% [17] and that of CNT intramolecular junction is only about 15% [14], while the rectification ratio in GR is about 270% and 350% for RGR and TGR, respectively. Our results demonstrate that GR rectifier has obvious advantage over the carbon nanotube based thermal rectifiers [14, 17]. Moreover, it is obvious that the rectification ratio of TGR is larger than that of RGR under the same temperature difference. This is consistent with the phenomena that carbon nanocone (geometric graded structure) has higher rectification ratio than the carbon nanotube *(n, 0)/(2n, 0)* intra-molecular junctions (which is two-segment structure) does. In these two-segment structures, such as nanotube junction and the RGR, although the phonon spectra in the two segments are quite different, they are homogeneous in each segment which will weaken the control on heat flux. In contrast, in the graded structures, the phonon spectra changes continuously, leads to more efficient control of heat flux.

In the following, we explore the size dependence of rectification of TGR, as the rectification of TGR is higher than that of RGR. First, the dependence of the rectification ratio on vertex angle, $\theta$, is explored. The length of TGR, L, is fixed at 3.4 nm, and $W_{top}$ is fixed at 0.4 nm. $\theta$ is changed from 26° to 78°, which corresponds to $W_{bot}$ from 1.7 to 5.9 nm. Figure 3(a) shows that the rectification increases with vertex angle, $\theta$, due to the increases of asymmetry. As illustrated in Figure 3(b), the heat fluxes, $J_+$ and $J_-$, of TGR increase as θ increases. The increase of $J_+$ is higher than the increase of $J_-$, which leads to the increase of the rectification ratio. Figure 3(a) shows the rectification ratio increase from 90% to 240% (here $|\Delta|=0.3$) as the increasing of $\theta$ from 26° to 60° and converges when $\theta \geq 60^o$.

Next, we study the impact of length on the rectification ratio. We change the value of L from 3.4 nm (30 layers) to 13.5 nm (120 layers), while the value of $\theta$ are fixed at 60°. The top end and the angle of TGR are fixed while the length L and bottom end $W_{bot}$ are changed. Figure 3(c) shows the dependence of rectification on the length. At short length, the small increase of length induces large reduction in rectification ratio. Contrast to the high dependence at the short length range, the rectification versus length curve is close to flat when length is longer than 5.1 nm. As illustrated in figure 3(d), the increase of $J_-$ is larger than that of $J_+$ as L increase from 3.4 nm to 5.1 nm. Both $J_+$ and $J_-$ are converged when L is longer than 5.1nm, which leads to the rectification ratio as a constant of 92%. This is different from previous results of one-dimensional structures. In the molecular junction typed CNT thermal rectifier [14], the rectification decreases quickly as device length is increased, because the role of the interface is suppressed in large system. However, in the GR rectifier proposed here, high rectification can be achieved in the practical length scale, adds the feasibility of constructing thermal rectifier with graphene.

In conclusion, we have demonstrated excellent thermal rectification efficiency in asymmetric graphene ribbons. It is shown that the graded geometric asymmetry is of remarkable benefit to increase the rectification ratio. The convergence of rectification ratio on the increase of length is of particular importance for practical thermal (phononics) application. Compared to the CNT [14] and CNC [17] rectifier, the rectification ratio in TGR is much higher. In contrary to the previous studies of thermal rectifiers which are constructed with one-dimensional structures, our results demonstrate the two-dimensional structures can also be used in thermal management.

This work is supported in part by an ARF grant, R-144-000-203-112, from the Ministry of Education of the Republic of Singapore, and Grant R-144-000-222-646 from National University of Singapore.

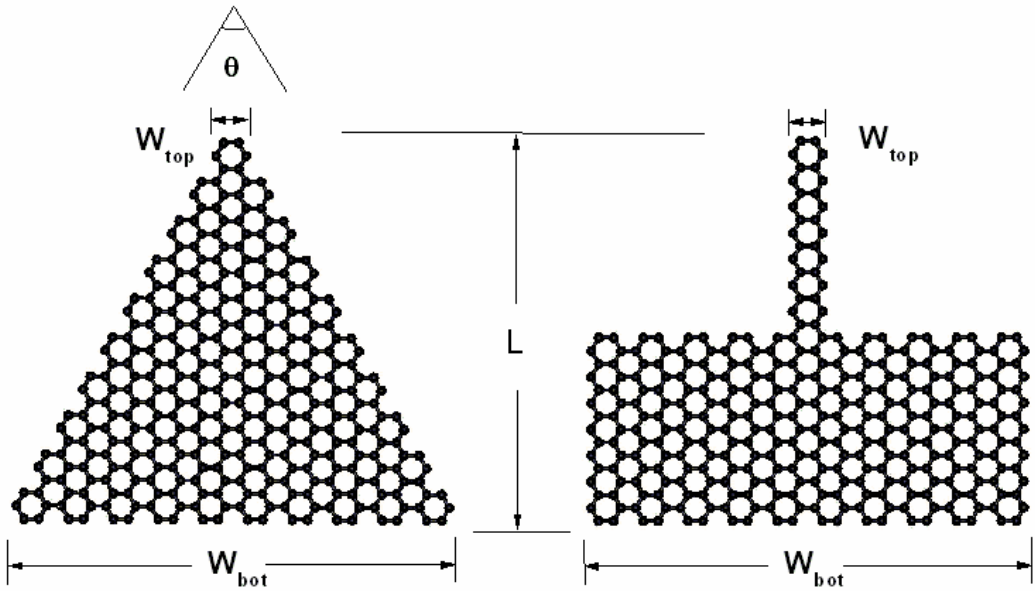

Figure 1. Schematic pictures of the two different graphene ribbons (GR) structures: trapezia shaped GR (TGR) and two-rectangular GRs with different width (RGR). The top layer is the first layer, with width of $W_{top}$, and the bottom layer is the *Nth* layer, with width of $W_{bot}$. Here *N* is the total number of layers. The length is *L*, while in RGR, each segment is with the same length of *L/2*.

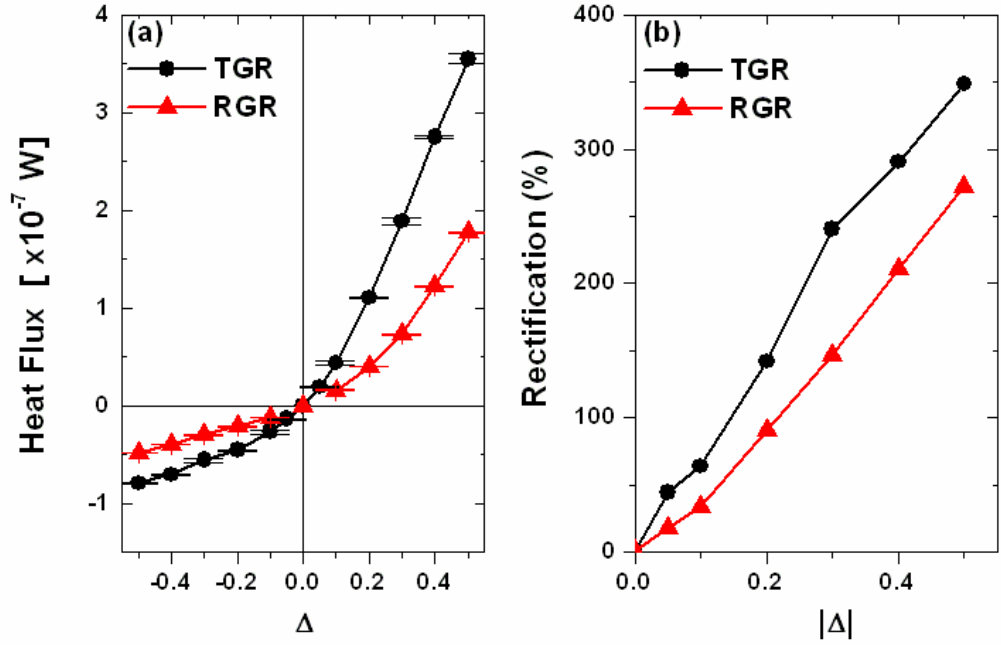

Figure 2. (Color on-line) (a) Heat flux J versus $\Delta$ for the TGR and RGR. (b) Rectifications versus $\Delta$ for the TGR and RGR. Here, L=3.4 nm (30 layers), $\theta = 60^o$, $W_{top}$ = 0.42 nm (2 atoms), $W_{bot}$ = 4.2 nm (20 atoms), and $T_0$ = 300 K.

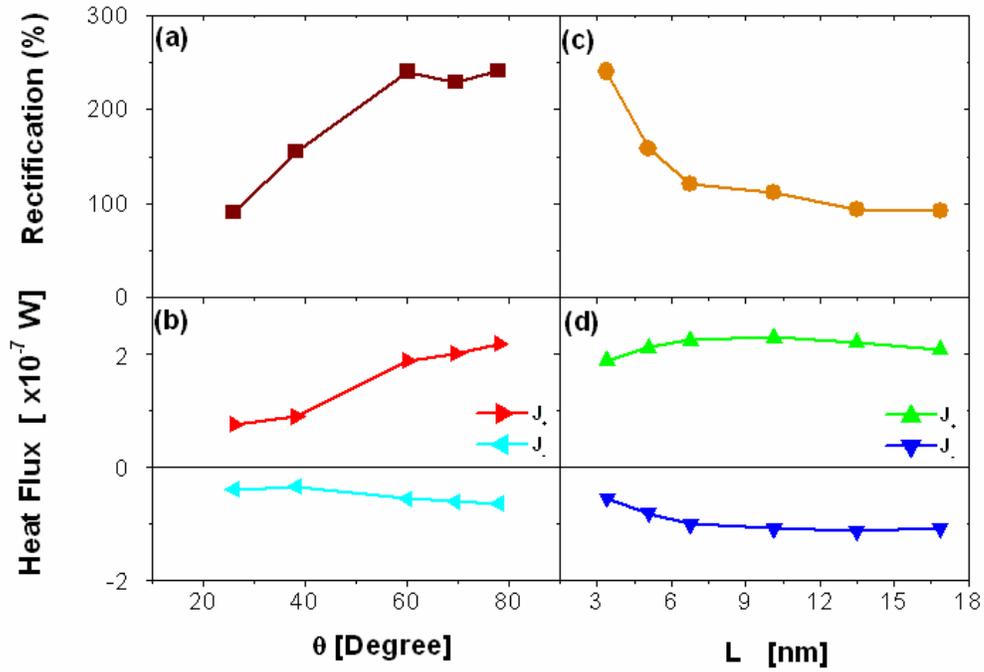

Figure 3. (Color on-line) (a) and (b) Rectifications and heat fluxes of TGR with different $\theta$, here L = 3.4 nm (30 layers), $W_{top}$ = 0.42 nm, $T_0$ = 300 K and $|\Delta| = 0.3$. (c) and (d) Rectifications and heat fluxes of TGR for different length, here $\theta = 60^o$, $W_{top}$ =0.42 nm, $T_0$ = 300 K and $|\Delta| = 0.3$.